# Anomalous Magnetoresistance in Centrosymmetric Skyrmion-Lattice Magnet Gd$_2$PdSi$_3$

H. Zhang[1*], Q. Huang[1], L. Hao[1], J. Yang[1], K. Noordhoek[1], S. Pandey[1], H. Zhou[1] and J. Liu[1*]

[1]Department of Physics and Astronomy, University of Tennessee, Knoxville, TN, USA, 37996

## Abstract

We performed a systematic study of the temperature- and field-dependence of magnetization and resistivity of Gd$_2$PdSi$_3$, which is a centrosymmetric skyrmion crystal. While the magnetization behavior is consistent with the reported phase diagram based on susceptibility, we show that a phase diagram can also be constructed based on the anomalous magnetoresistance with one-to-one correspondence among all the features. In addition, the crossover boundary into the field-induced ferromagnetic state is also identified. Our results suggest that the ferromagnetic spin fluctuations above the Néel temperature play a key role in the high sensitivity of the resistivity anomalies to magnetic field, pointing to the rich interplay of different magnetic correlations at zero and finite wave vectors underlying the skyrmion lattice in this frustrated itinerant magnet.

*corresponding author

## Introduction

Magnetic skyrmions, named after the British Physicist Tony Skyrme, are used to describe vortex-like complex spin textures in magnetic materials [1-3]. skyrmion textures can arise from a variety of emergent phenomena, including quantum Hall effect [4], liquid crystals [5] and spinor Bose condensates [6]. One of the most celebrated properties of the skyrmion is that the 'winding number' of its spin texture is a topological integer which stays invariant under smooth fluctuations of the spin configuration [1-3]. Therefore, skyrmions can potentially be used as the next-generation information carrier with lower energy cost [7,8]. The first observations of skyrmion crystals were reported in non-centrosymmetric materials, where the competition between Heisenberg and Dzyaloshinskii-Moriya (DM) interactions leads to twisting among neighboring spins [9-13].

Shortly after the discovery of skyrmions non-centrosymmetric materials, it was recognized that skyrmions can also emerge in centrosymmetric magnets [14-20], which was very recently observed by Kurumaji et al. in geometrically frustrated magnet $Gd_2PdSi_3$ [21]. $Gd_2PdSi_3$ crystalizes in the hexagonal space group P6/mmm [22], where the triangular Gd layers are separated by Pd/Si layers of honeycomb structure as shown in figure 1(a). It was reported to exhibit two metamagnetic transitions below Néel temperature $T_N$ = 20 K with an unusual Hall signal in-between [23]. Interestingly, the phase in between the metamagnetic transitions turns out to be a skyrmion crystal with a giant topological Hall effect [21] produced by a much shorter period of the spin modulation in comparison with the conventional skyrmion materials [24,25]. The high skyrmion density results in a skyrmion lattice described by a superposition of three helical orders with incommensurate wave vectors. This novel phenomenon has sparked the interest of the community for understanding its underlying physical mechanisms. As an intermetallic magnet, understanding the interplay between the itinerant electrons and the local Gd moments is believed to be vital.

In this work, we performed systematic study of the magnetization and resistivity of $Gd_2PdSi_3$ under a range of temperatures and magnetic fields. We show that the resistivity of the system also exhibits anomalies that are highly sensitive to the spin fluctuations near at the phase boundaries, especially the ferromagnetic fluctuations above the Néel temperature. As a result, the magnetic field is able to drastically change the resistivity minimum from U-shape to V-shape. A one-to-one correspondence is established between the features of resistivity and magnetization at different temperatures and fields, enabling the construction of a resistivity phase diagram that mirrors the magnetic one.

**Experimental Details**

A single crystal of $Gd_2PdSi_3$ was grown by utilizing the floating zone method. Details can be found in refs. [26,27]. Lab x-ray diffraction characterizations of the samples were done with a Panalytical Empyrean x-ray diffractometer. Magnetization was measured with a Quantum Design superconducting quantum interference device (SQUID) magnetometer. Sample environment for the transport measurement was controlled by a Quantum Design Physics Properties Measurement System (PPMS). Hall and resistivity measurements were performed with a standard four-probe method with the lock-in technique.

**Results and Discussion**

As we explained above, $Gd_2PdSi_3$ has been recently identified as a host for a magnetic field-induced skyrmion lattice with a giant topological Hall signal. Below the Néel temperature, a magnetic field perpendicular to the *ab*-plane stabilizes the skyrmion lattice between two metamagnetic transitions [21,23]. Figure 1(b) shows the DC magnetization and Hall resistivity measured at 2 K, with an external magnetic field applied along the [001] direction. The Hall measurement can be interpreted as a superposition of a small ordinary Hall effect and a topological Hall signal with a maximum amplitude about 2.5 µOhm·cm between ~0.5 T and ~1.1 T. These field values coincide with the kinks of the $M(H)$ curve, which correspond to two peaks in the field derivative of the magnetization $\chi(H) = dM/dH$. As shown in figure 1(c), by tracking the evolution of $\chi$ as a function of $H$ at different temperatures, we successfully reproduced the temperature-field phase diagram reported in ref. [21]. From the color scale of $\chi$, one can readily see the two metamagnetic phase boundaries in yellow at low fields, separating the so-called IC-1 phase at zero field, the skyrmion lattice phase at intermediate fields, and the so-called IC-2 phase below the saturation field. These are incommensurate magnetic configurations characterized by multiple- or single-ordering wave vectors $\boldsymbol{Q}_\nu$ ($\nu = 1,2,3$) that differ by 120° rotations about the *c*-axis. Upon further increasing the field, $\chi$ evolves from the light blue region of the IC-2 phase to a dark blue region. The boundary of these two regions is not associated with a peak in $\chi(H)$ but with a small value of $\chi$ that is weakly dependent on field and temperature. According to the reported AC susceptibility [21], this boundary signals the transition to the paramagnetic phase that is adiabatically connected with the fully polarized state at $T = 0$. We have verified this phase evolution by measuring the $M(T)$ curves shown in Figure 2 (a), which displays a sharp peak typical of a Néel transition at zero field. This peak shifts to lower temperatures and becomes a broad hump as the field increases. The positions of this maximum are marked on the phase diagram as black diamonds, and they indeed coincide with the IC-2-to-paramagnetic phase boundary.

The drastic change of shape of $M(T)$ near the transition contains rich information on the competition between the underlying magnetic interactions. In particular, the $M(T)$ curve is significantly flattened below the transition at high fields. The overall temperature dependence is reminiscent of a ferromagnet under applied field. In fact, $Gd_2PdSi_3$ has been known to show a positive Curie-Weiss temperature $\Theta$ [23]. Figure 2 (b) shows the inverse of susceptibility between 5 K and 80 K measured under a small field of 0.1 T, which follows the Curie-Weiss law at $T > 40$ K. A linear regression reveals a Curie-Weiss temperature $\Theta = 27.4$ K, which is consistent with the previous report [23]. The Gd moments clearly experience an overall ferromagnetic (FM) correlation. By examining the $M(H)$ curves up to 3 T above the Néel transition, we readily observed a nonlinear behavior characteristic of a Brillouin function [28]. For example, as shown in figure 2(c), the $M(H)$ curve at 40 K can be well-described by the Brillouin function. The same figure shows that this Brillouin function-like behavior is enhanced upon further cooling toward the Néel transition. Indeed, the $M(H)$ curve at 30 K clearly begins to show saturating behavior at fields above ~1.5 T, implying that the field is enforcing a parallel alignment of the Gd moments assisted by the FM correlations. This observation explains the high-field $M(T)$ curves as a result of the suppression of the incommensurate state and a field-induced FM state emerging above the Néel transition. We extracted the crossover temperature scale $T^*$ from the inflection point of the $M(T)$ curve, i.e., $(\frac{dM}{dT})_{min}$, and overlaid the resulting $T^*(H)$ curve on the phase diagram in figure 1(c). One can see that $T^*$ is slightly above $T_N$ at small fields and it increases with the field. The interval $T^*$-$T_N$ increases with the field and matches the dark blue region of $\chi$ above 1 T as the magnetization approaches the saturated value. The combined results suggest that FM interactions, which produce dominant short-range FM correlations at $T > T^*$ compete against antiferromagnetic interactions that become relevant below $T^*$ and lead to the incommensurate spin-state that is stabilized below $T_N$. In other words, as expected for centrosymmetric materials, the spiral ordering of $Gd_2PdSi_3$ seems to arise from the frustration produced by competition between FM and AFM interactions between magnetic ions separated by different distances. We note that this real space picture of the exchange interaction is consistent with the momentum space dependence of the effective exchange interaction that was recently derived from first principle calculations [29].

As the magnetic interactions are mediated by itinerant electrons in $Gd_2PdSi_3$, the different magnetic regimes have a strong impact on the electronic transport: within the Born approximation, the electron-spin scattering is controlled by the magnetic structure factor $S(\bm{k})$. We then performed systematic resistivity measurements under various temperatures and magnetic fields, which are displayed in figure 3. In agreement with previous reports [23, 30], the resistivity between 2 K and 300 K exhibits an overall metallic behavior [see inset of figure 3(a)]. There is an upturn below 40 K until reaching ~ 20 K [30], below which the metallic behavior is recovered. This anomalous behavior near $T_N$ is characteristic of critical spin fluctuations in AFM metals described by a spin-spin correlation function at finite wave vectors [31,32].

Close examination of the resistivity upturn reveals that it includes a fine structure, which turns out to have a strong field-dependence. In particular, there is a kink in between the minimum and the maximum that is shown in figure 3(a). Upon applying an external magnetic field $\bm{H} \parallel [001]$, the upturn rapidly evolves into a downturn for $H > 1\,T$. A resistivity downturn is usually associated with the development of critical $Q=0$ magnetic fluctuations in FM metals

[31]. Indeed, the temperature of the downturn kink, tracked by the red dashed line, increases with the field similarly to the crossover temperature $T^*(H)$ obtained from the results of the magnetization. Meanwhile, the kink tracked by the blue dashed line becomes steeper and shifts to lower temperatures as the field increases, while the Néel transition is suppressed as indicated by the shift of the resistivity maximum to lower temperatures. The combination of these behaviors results in a drastic change in the shape of the resistivity curve: the resistivity upturn evolves from a U-shape minimum to a sharp V-shape minimum due to the field-enhanced FM correlations.

The overall resistivity reduction reaches ~30% within our field range indicating a negative MR. However, this behavior is qualitatively different at small fields and below $T_N$, where the MR is positive, as it is shown in figure 3(a). A closer look at the low-field data is displayed in figure 3(b). While the maximum is shifted to lower temperatures, the resistivity increases for low fields (positive MR). Figure 3(c) shows several characteristic MR curves at selected temperatures. Above $T_N$, the MR curve is always a nonlinear negative response with an increasing magnitude at lower temperatures. There is also a sign change of the curvature when approaching the resistivity minimum region, which can be related to the dramatic field-induced U-shape-to-V-shape change. When the temperature is below $T_N$, the resistivity first increases with the magnetic field, and then turns into a weak negative MR behavior at a critical field, which appears as a plateau-like feature. At another higher critical field, the negative MR response suddenly increases. The two critical fields coincide with the two metamagnetic transitions, and the plateau-like MR is associated with the skyrmion lattice phase. As the field is further increased, the negative MR response slows down and shows another slope change, which can be seen in the MR at 10 K (red solid line in figure 3(c)) and can be related to the fact that the system eventually enters the field-induced FM phase.

Given the strong correspondence between the phase evolution and the resistivity, we constructed a color map of the MR data, which is displayed in figure 3(d) and clearly mirrors the field-temperature phase diagram by susceptibility (figure 1(c)). To quantitatively determine the phase boundaries on the resistivity phase diagram, we compared the temperature derivative of resistivity $d\rho/dT$ with $M(T)$ and $dM/dT$. Comparisons are demonstrated in figure 4 at selected temperatures and reveal the one-to-one correspondence between resistivity and magnetization features. We found that the valley in $d\rho/dT$ under different magnetic fields always has the same temperature as the peak in the $M(T)$ curve, as pointed out with the red dashed line for the 0.5 T result in figure 4 (a) and (b). This temperature of the maximized negative slope of $\rho$ corresponds to the Néel transition, consistent with the expectation from spin fluctuations in AFM metals [31,32]. On the other hand, a peak in $d\rho/dT$ is found to emerge with magnetic field at the same temperature as the valley of $dM/dT$, the magnitude of which is also enhanced by the field as shown in figure 4 (a) and (c) (red dash-dot line showing the case for $H$ = 0.5 T). This feature characterizes the temperature of the maximized positive slope of the V-shape minimum, corresponding to the crossover temperature $T^*$. We mark both of these temperatures extracted from resistivity on the phase diagram in figure 3(d) to indicate the phase boundaries. One can see from the color scale of figure 3(d) that the region in-between covers the V-shape resistivity minimum and the large MR within the field-induced FM state. Finally, the phase boundary of the skyrmion lattice phase is indicated by the critical fields of the plateau-like MR feature, which surround a relatively uniform region of positive MR highlighted in red on figure 3(d). Therefore, the longitudinal resistivity of $Gd_2PdSi_3$ is also a good indicator of its emerging skyrmion lattice and the underlying competing magnetic interactions.

From a microscopic point of view, a U-shape resistivity minimum is commonly seen in itinerant systems with local magnetic moments and is often attributed to the Kondo effect [33]. However, an increasing number of studies suggest that a RKKY-interaction-induced classic spin ice state can also stabilize such a resistivity upturn [34-37]. A recent work [38,39] suggests the finite-$Q$ spin-spin correlation driven by the RKKY interactions can be responsible for such an upturn in $Gd_2PdSi_3$. Its combination with the inherent FM correlation leads to the dramatic shape change.

In summary, we have successfully synthesized $Gd_2PdSi_3$ single crystals and performed systematic magnetization and electrical resistivity studies. The resistivity upturn and negative MR above the magnetic transition temperature could be originated by competing FM and AFM interactions. We have also observed a strong correlation between the charge and spin degrees of freedom near all phase boundaries, including the skyrmion lattice. Longitudinal resistivity can thus be used as a good indicator of the magnetic skyrmion phase and the nearby competing phases along with the ac susceptibility/Hall resistivity. This observation highlights the critical role of the charge degrees of freedom in the formation of the complex spin configurations of this centrosymmetric frustrated magnet. In preparation of this manuscript, we notice an independent report that suggests a so-called "depinning phase" within the IC-2 phase in the high field region [40]. However, the corresponding signatures are not observed our magnetization and resistivity data.

**Acknowledgement**

We thank Z. Wang and C. D. Batista for useful discussions. This work is support by U.S. Department of Energy under grant No. DE-SC0020254. J.L. and H.D.Z. acknowledge support from the Organized Research Unit Program at the University of Tennessee. Use of the PPMS and SQUID is supported by the Electromagnetic Property (EMP) Lab Core Facility at the University of Tennessee.

**Figure 1**

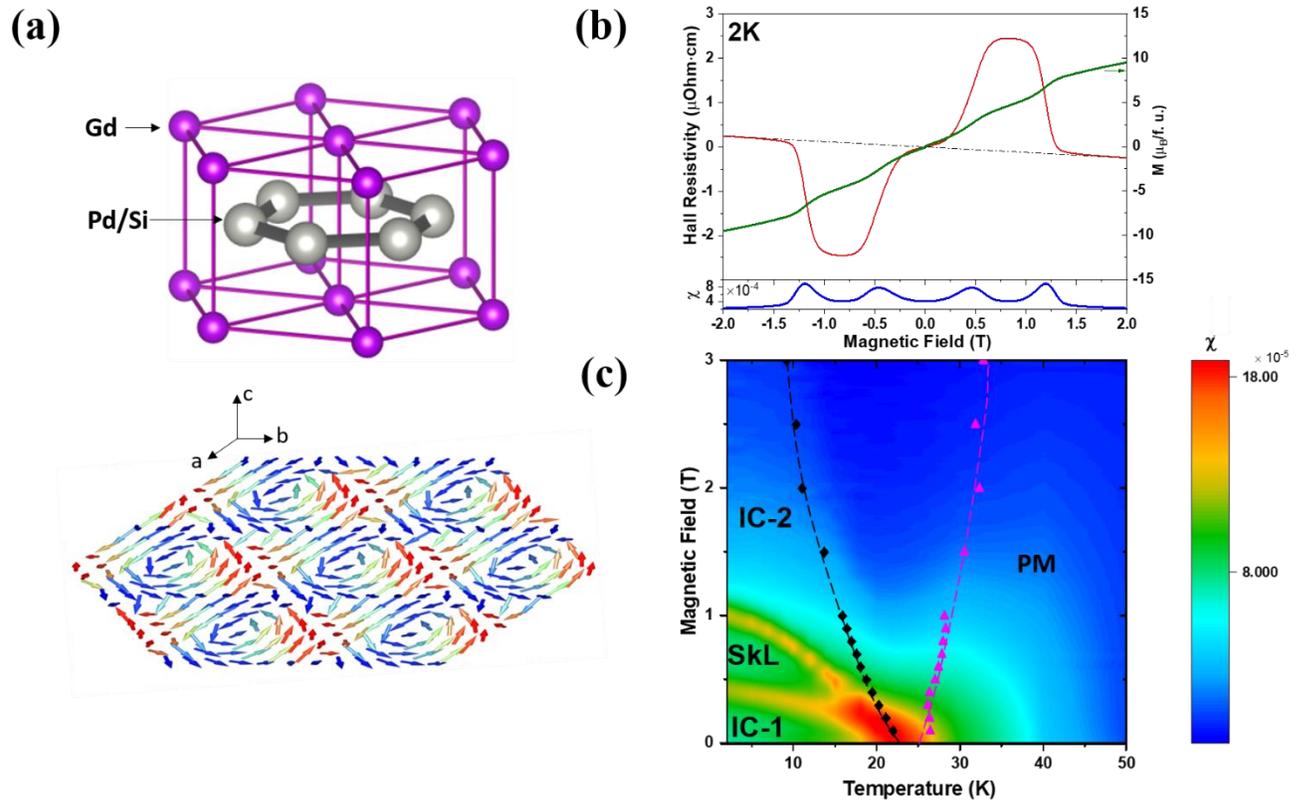

**Figure 1** (a) (top) crystal structure of $Gd_2PdSi_3$. The triangular Gd layers are separated by Pd/Si layers of Honeycomb structure. (bottom) skyrmion spin texture formed on the magnetic Gd atom sites. (b) Magnetization (green) and the Hall resistivity (red) of $Gd_2PdSi_3$ measured at 2K. (c) H-T phase diagram of $Gd_2PdSi_3$ generated from magnetic susceptibility. SkL represents the skyrmion phase while PM denotes the paramagnetic phases. IC-1 and IC-2 are two incommensurate spin-state phases.

**Figure 2**

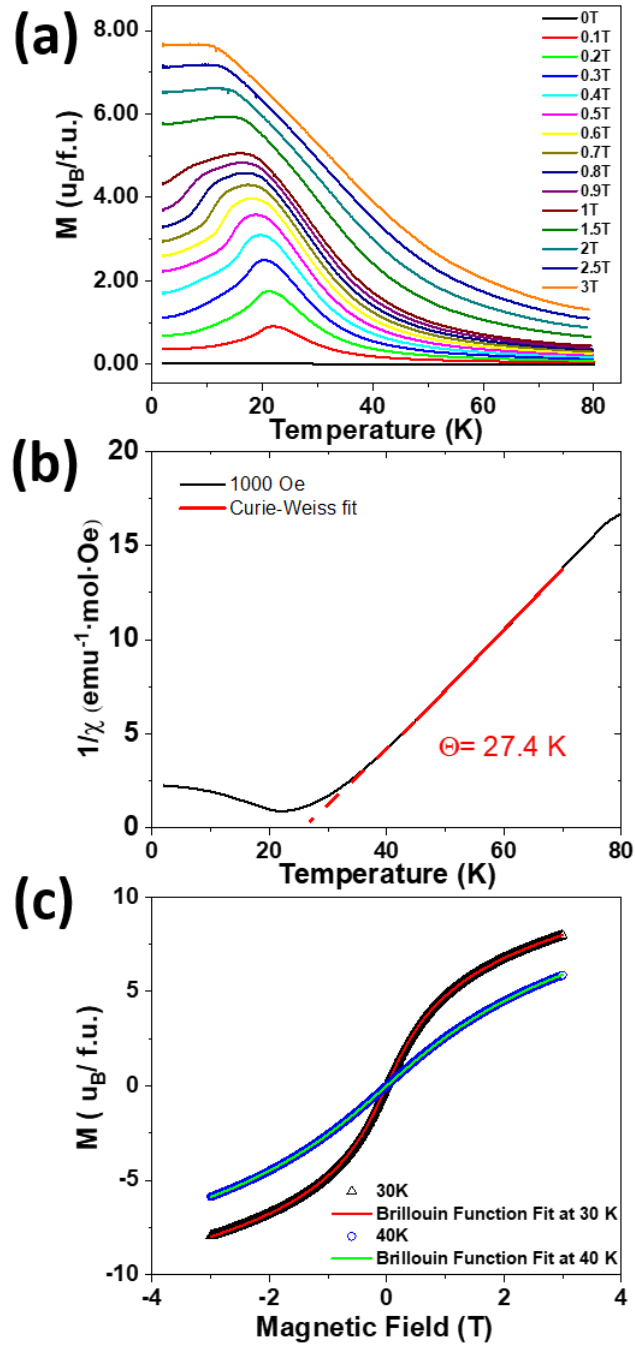

**Figure 2** (a) Magnetization *vs.* temperature at different magnetic fields. (b) 1/χ (black) and the Curie-Weiss fit (red) at 1000 Oe of $Gd_2PdSi_3$. The Curie Temperature $T_c$ is extracted to be Θ = 27.4 K, suggesting ferromagnetic interaction. (c) Magnetization vs. magnetic field between -3 T and 3 T at 30 K and 40 K (symbols). The shape highly suggests a ferromagnetic nature of the system, and can be described well with a Brillouin function (solid lines)

**Figure 3**

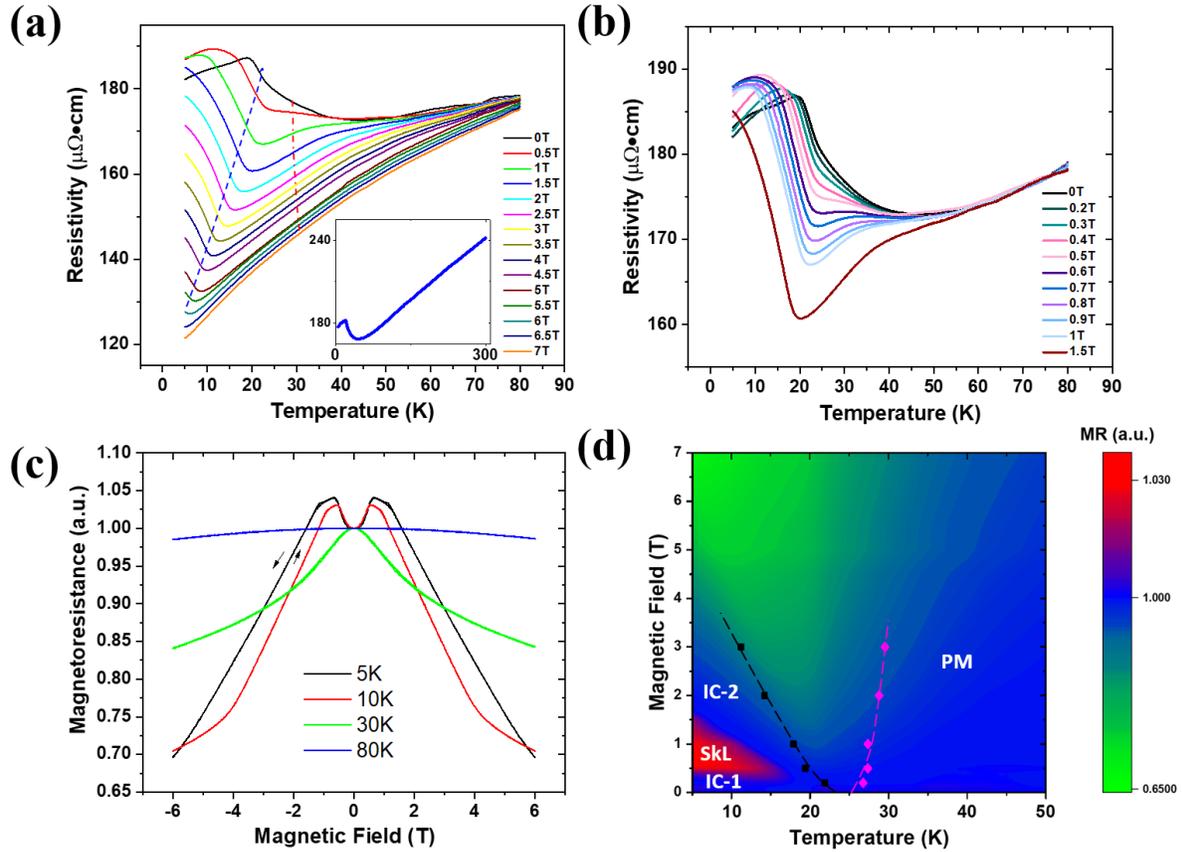

**Figure 3** (a) Resistivity of $Gd_2PdSi_3$ measured between 5 K and 80 K under external magnetic field between 0 T and 7 T, inset shows the resistivity between 5 K and 30 K under ambient magnetic field; The blue and red dashed lines track the evolution of two kinks with increasing magnetic field. (b) Resistivity of $Gd_2PdSi_3$ measured between 5 K and 80 K under external magnetic field between 0 T and 1.5 T (c) The measured Magnetoresistance (MR) at several selected temperatures (d) H-T phase diagram acquired from MR curves.

**Figure 4**

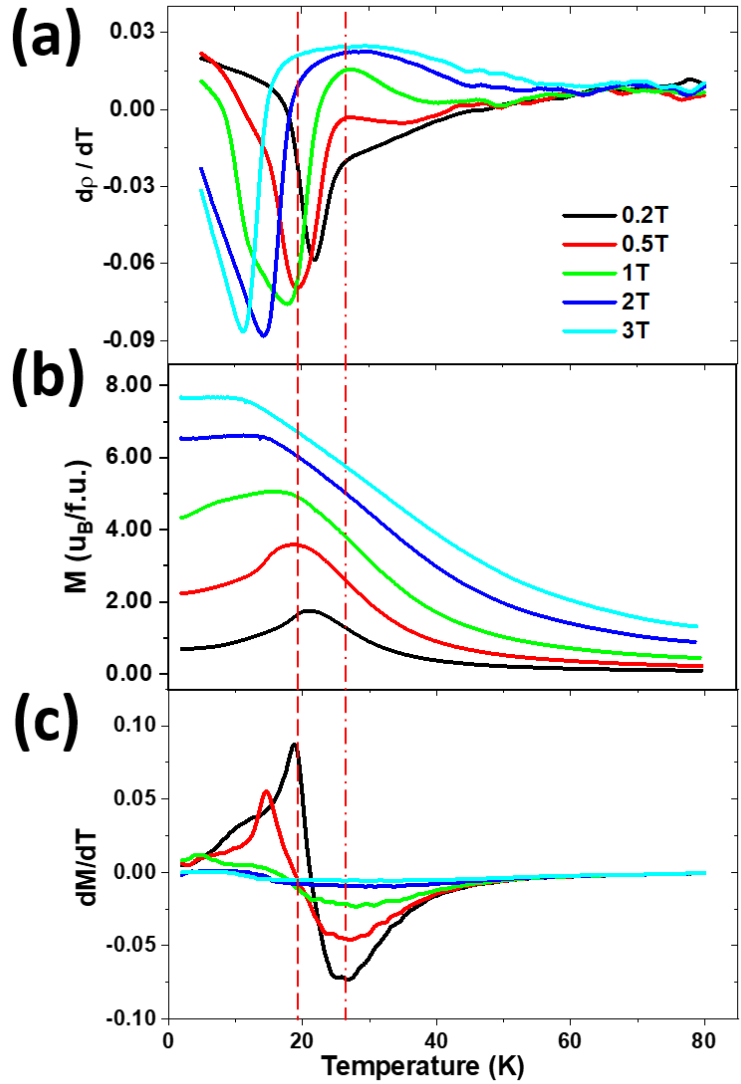

**Figure 4** (a) resistivity derivative $d\rho/dT$, (b) magneticzation and (c) temperature derivative of the magnetization measured between 5K and 80K, under selected magnetic field between 0 T and 3 T. The red dashed line for H = 0.5 Tshows the valley in $d\rho/dT$ under different magnetic fields always has the same temperature as the peak in the $M(T)$ curve; whereas the dash-dot line suggests a peak in $d\rho/dT$ always emerge with magnetic field at the same temperature as the valley of $dM/dT$.